# Three-dimensional atomic interface between metal and oxide in Zr-ZrO$_2$ nanoparticles

Yao Zhang[1,5], Zezhou Li[1,5], Xing Tong[2,5], Zhiheng Xie[1], Siwei Huang[1], Yue-E Zhang[2,3], Hai-Bo Ke[2]*, Wei-Hua Wang[2,4], Jihan Zhou[1]*

[1]Beijing National Laboratory for Molecular Sciences, Center for Integrated Spectroscopy, College of Chemistry and Molecular Engineering, Peking University; Beijing, 100871, China.
[2]Songshan Lake Materials Laboratory, Dongguan 523808, China.
[3]College of Physics, Liaoning University, Shenyang 110036, China.
[4]Institute of Physics, Chinese Academy of Sciences, Beijing 100190, China.
[5]These authors contributed equally to this work.
*Correspondence and requests for materials should be addressed to H.-B. K. (email: kehaibo@sslab.org.cn) and J. Z. (email: jhzhou@pku.edu.cn)

**Abstract**

Metal-oxide interfaces with poor coherency have unique properties comparing to the bulk materials and offer broad applications in the fields of heterogeneous catalysis, battery, and electronics. However, current understanding of the three-dimensional (3D) atomic metal-oxide interfaces remains limited because of their inherent structural complexity and limitations of conventional two-dimensional imaging techniques. Here, we determine the 3D atomic structure of metal-oxide interfaces in zirconium-zirconia nanoparticles using atomic-resolution electron tomography. We quantitatively analyze the atomic concentration and the degree of oxidation, and find the coherency and translational symmetry of the interfaces are broken. Moreover, we observe porous structures such as Zr vacancies and nano-pores and investigate their distribution. Our findings provide a clear 3D atomic picture of metal-oxide interface with direct experimental evidence. We anticipate this work could encourage future studies on fundamental problems of oxides such as interfacial structures in semiconductor and atomic motion during oxidation process.

## Introduction

The oxidation is ubiquitous and involved in many processes on a daily basis. Most metals spontaneously form an oxidation layer on their surfaces. The metal-oxide interface plays a critical role in broad applications ranging from heterocatalysis[1,2], batteries[3,4], and electronics[5,6]. The thermodynamics and kinetics of oxidation process have been extensively studied over the years[7–9]. Plenty of research has been focused on the metal's work function[10], the transport of metal or oxygen species[11–13], and the rate of oxidation[14]. A number of theories have been proposed upon these studies to understand the oxidation behavior. For example, Kirkendall effect is used to explain the formation of oxidized pores[15–17]; Wagner proposed the law of oxidation kinetics in which the oxidation rate is controlled by the transport of ion under electrochemical potential gradient based on several assumptions[12,14,18]. However, owing to the lack of direct observation of the metal-oxide interface at nanoscale or atomic scale, most of the classical oxidation theories have been limited to the macroscopic scale. Properties of the interface including catalytic activity, phonon dispersion and electron transportation are strongly related to the local atomic arrangements of the metal-oxide interface such as coordination numbers and atomic bond lengths[19–24]. It is therefore essential to determine the three-dimensional (3D) atomic arrangements and understand the detailed oxidation structure of metal.

With the recent development in aberration corrected transmission electron microscopy (TEM), local structures of metal-oxide interfaces such as $Cu-Cu_2O$[25,26], $Ag-Ag_2O$[27], and $Ni-NiO$[28] have been studied at nanoscale or atomic scale; several of them were probed *in situ* by atomic resolution imaging and theoretical simulation[25–29]. Semi-coherent and incoherent interfaces between metal and oxide have been observed at sub-angstrom resolution from two dimensional (2D) projections[26]. Luo et al. discovered the periodic dislocation in $Cu-Cu_2O$ semi-coherent interface, suggesting the mechanism of strain release by defects between metal and oxide[25]. Zhu et al. tracked the formation of voids in Ni-NiO nanoparticles at nanoscale, identifying a two-stage oxidation mechanism including early-stage nucleation and then the Wagner oxidation[28]. However, since the oxidized interfaces are usually non-epitaxial and inherently disordered due to the lattice mismatch, the atomic arrangements between some metal and its oxidation layer cannot be clearly elucidated using high resolution TEM or crystallography. Conventional 3D characterization methods such as atom probe tomography[30,31], electron tomography[28,32–35], and depth sectioning[36,37] have been used to study the 3D morphological structures of the oxidized interfaces, and these techniques could overcome the limitation of single images which only provide the projected information of the 3D structures in 2D. However, the resolution of these techniques has limited to nanometer scale. Thus, determining the 3D atomic arrangements of the metal-oxide interface remains a major challenge. Although it remains notoriously difficult to imaging and identify each of the oxygen atoms of oxides in 3D, especially in high-angle annular dark-field scanning transmission electron microscopy (HAADF-STEM) mode, atomic resolution electron tomography (AET), which is an effective tool for determining the 3D atomic structure of nanomaterials[32–35], can in principle resolve the positions of heavy metal atoms in oxides and therefore give important structural information on this long-standing problem.

Here using $Zr-ZrO_2$ as a model system, we determine the 3D atomic structure of the metal-oxide interface using AET. We choose $Zr-ZrO_2$ for two reasons, first, Zr can form oxide spontaneously in air and the oxidation process is moderate[10]; second, the Zr-O bonding is extremely strong among all the common metal oxides and $ZrO_2$ has extreme chemical stability, the Zr-ZrO$_2$ interface can maintain its atomic structures after electron illumination at a dose rate of 6×10$^5$ e·Å$^{-2}$, which is essential for electron tomography experiment. By determining all the Zr atomic positions in Zr-ZrO$_2$ nanoparticles (NPs), we obtained the 3D atomic structure of a partially oxidized Zr NP; it has an uncommon face-centered cubic (FCC) Zr metal crystal nucleus as the core and amorphous/crystalline ZrO$_2$ as the shell. We observed the atomic packing heterogeneity and numbers of Zr vacancies and small nano-cracks in the oxidation shell. Instead of forming a coherent interface, most of the atoms at the Zr-ZrO$_2$ interfaces connect with each other semi-coherently or incoherently. The degree of oxidation is decreasing while Zr packing density is increasing from the oxide surface to the metal core. We discovered a bidirectional distortion including bending and twisting at the semi-coherent metal-oxide interface. Moreover, we identify numbers of voids in the oxides including Zr vacancies, nano-pores and large pores; the oxidation process is related to the distribution of the voids. These findings expand our understanding of the atomic structures of metal-oxide interfaces with poor coherency, encourage future studies on oxidation process at 3D atomic resolution, and further inspire the designing and modeling of atomic metal-oxide interface in surface engineering, heterogeneous catalysis and semiconductor.

**Results**

**Atomic structures of Zr-ZrO$_2$ nanoparticles in 3D**

NPs made of different monatomic metals with both disordered and crystalline structures can be achieved using fast-cooling vitrification process[38,39]. Zr NPs were synthesized using pulse laser ablation of pure Zr target (purity > 99.95%) in ethanol (Methods). By naturally oxidizing the freshly prepared Zr NPs in air, we obtained Zr-ZrO$_2$ NPs at different stages of the oxidation process. Some of the Zr-ZrO$_2$ NPs have an oxidized shell and a metal core (Supplementary Fig. 1). To confirm the oxidation, we used high resolution HAADF-STEM, energy dispersive spectroscopy (EDS) and electron energy loss spectroscopy (EELS) to characterize the Zr-ZrO$_2$ NPs (Supplementary Fig. 2 and Supplementary Fig. 3, respectively); it is notable that the edges of the NPs have a high degree of oxidation since the oxygen signal of EELS at the edge are stronger while HAADF-STEM intensities are weaker.

We resolved the 3D atomic structures of all Zr atoms in several Zr-ZrO$_2$ NPs using AET. In short, tomography tilt series (Supplementary Figs. 4-6) were acquired from three Zr-ZrO$_2$ NPs at different stages of the oxidation process using an aberration-corrected TEM in HAADF-STEM mode (Supplementary Table 1). After imaging processing including denoising, background subtracting and alignment (Methods), the tilt series were reconstructed using algorithm described elsewhere[33–35]. The 3D atomic coordinates of all Zr were traced from the computed reconstructions (Methods). We chose a partially oxidized Zr-ZrO$_2$ (named Zr1) as our main interest to elucidate the metal-oxide interfaces. The other two particles are fully oxidized without obvious metal core (named as Zr2 and Zr3; Supplementary Fig. 7). Fig. 1a and Supplementary Movie 1 show the experimental 3D atomic model of Zr1, showing there are ordered crystalline grains and disordered structures in the particle. We calculated the normalized bond orientational order (BOO) parameters for all the atoms to quantify the disorder (Fig. 1b and Methods); about 32 % of all the Zr atoms are disordered. The particle has complicated phases, composing of a central metal grain, crystalline oxide grains (c-ZrO$_2$) and an amorphous oxide phase (a-ZrO$_2$) (Fig. 1d and Supplementary Movie 2). We calculated the Zr-Zr partial pair distribution functions (PDFs) of the atoms in c-ZrO$_2$ and a-ZrO$_2$ separately (Methods). The c-ZrO$_2$ show a well-matched cubic phase ZrO$_2$ structure instead of monoclinic phase ZrO$_2$ in both Zr1 and Zr2 (Fig. 1c). The PDFs of the a-ZrO$_2$ atoms in all three NPs exhibit similar shape; and the first peak position is located at 3.45 Å, which is close to the first peak position of monoclinic phase ZrO$_2$ (Fig. 1c). All the positions of main peaks and valleys in our Zr-Zr PDFs obtained from the atomic coordinates of our a-ZrO$_2$ structures agree with those obtained from synchrotron X-ray diffraction[40]. The most populated Zr-Zr bond lengths in c-ZrO$_2$ and a-ZrO$_2$ are 3.6 Å and 3.45 Å, respectively (Supplementary Fig. 8). Interestingly, there is a small Zr metal core inside Zr1, confirmed by polyhedron template matching[41] and atomic concentration analysis (Methods). Fig. 1e shows the atomic structure of the pure Zr metal core viewing from ⟨110⟩ direction. The metal core has a distorted FCC structure with averaged Zr-Zr bond length being 3.3 Å, slightly longer than the standard value in Zr metal (3.2 Å). These observations are different from the bulk behavior, where Zr typically forms hexagonal close-packed (HCP) or body-centered cubic (BCC) structures; and usually forms monoclinic phase ZrO$_2$ after natural oxidation[42]. This discrepancy highlights the distinctive behavior of materials at the nanoscale.

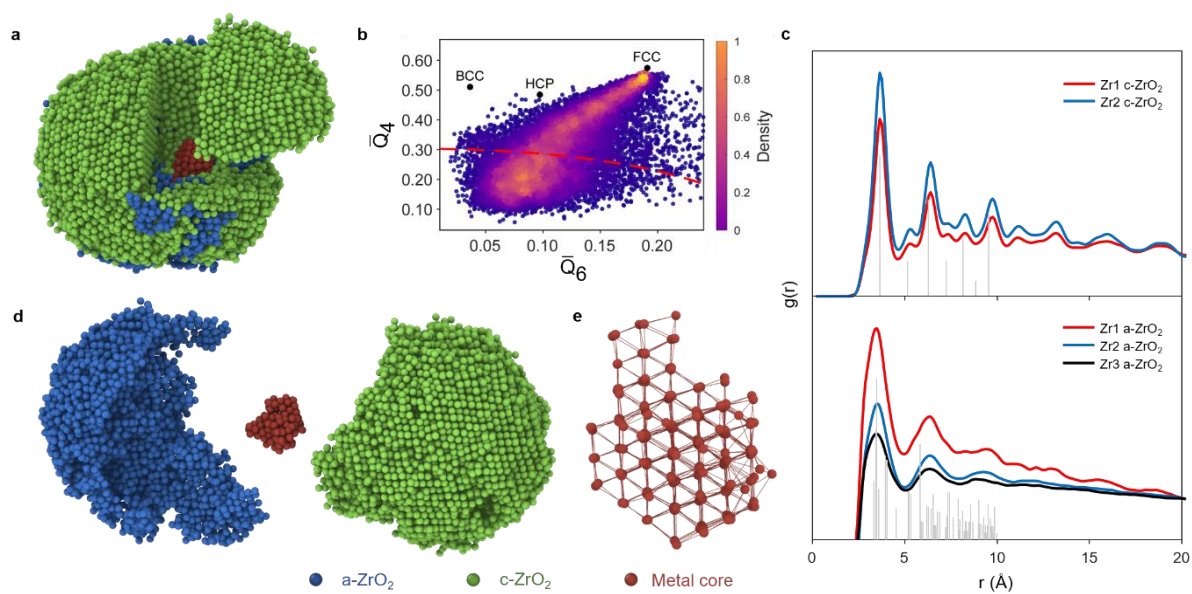

**Fig. 1 | Atomic structures of Zr-ZrO$_2$ nanoparticles in 3D. a**, Experimental 3D atomic model of the Zr1 NP with a-ZrO$_2$ in blue, c-ZrO$_2$ in green and metal core in red. **b**, Normalized BOO parameters of all atoms. The red dashed curve is a criterion to distinguish the disordered atoms (32% in total, atoms below the curve) and ordered atoms (68% in total, atoms above the curve). The standard BCC, HCP and FCC parameters are marked as black dots for reference. **c**, Zr-Zr PDFs of the c-ZrO$_2$ (top panel) and a-ZrO$_2$ (bottom panel), with Zr1 in red, Zr2 in blue and Zr3 in black. The gray peaks show the peak positions of the standard PDF of cubic phase ZrO$_2$ (top panel) and monoclinic phase ZrO$_2$ (bottom panel) for comparison. **d**, The NP consists of a-ZrO$_2$, c-ZrO$_2$ and a metal core grain. **e**, Magnified atomic structure of the pure Zr metal core viewing from ⟨110⟩ direction.

**Atomic concentration and the degree of oxidation**

To compare the local atomic packing density of Zr in all phases, we obtained the compactness of the NP by determining the Zr atomic concentration ($\rho_N$) of all the regions present in Zr1 NP

(Methods). Fig. 2a shows the 3D $\rho_N$ distribution of Zr1. The low packing density regions are not related to any voids in the NP as we exclude all the voids from consideration when performing calculation (Methods). The averaged $\rho_N$ of metal core is $3.85 \times 10^{-2}$ Å$^{-3}$ (Fig. 2c), close to the $\rho_N$ of ideal close-packed metallic Zr ($3.9 \times 10^{-2}$ Å$^{-3}$). The $\rho_N$ of oxides (both c-ZrO$_2$ and a-ZrO$_2$ phases) are significantly lower than that of pure metal, being $2.92 \times 10^{-2}$ Å$^{-3}$ and $2.89 \times 10^{-2}$ Å$^{-3}$, respectively. They are comparable to the $\rho_N$ of ideal c-phase ZrO$_2$ ($3.0 \times 10^{-2}$ Å$^{-3}$). We also observed 3D local $\rho_N$ heterogeneity in the oxides particularly distributed around the metal-oxide interfaces. Fig. 2d shows the $\rho_N$ distribution as a function of the distance from the surface of metal core (metal to c-ZrO$_2$). The gradually decrease in $\rho_N$ suggests the metal-oxides interfaces are atomically smooth interface. The packing density gradient is attributed to the gradual change of the degree of oxidation of the Zr metal. Our PDFs and Zr-Zr bond length analysis suggest that c-ZrO$_2$ is c-phase, and a-ZrO$_2$ mainly forms the tetrahedral structure locally;[40] oxygen should locate in tetrahedral sites in both phases (Supplementary Fig. 9). Next, we quantified the degree of oxidation by geometrically filling oxygen into the tetrahedral sites (Methods). Since EDS and EELS measurements in other similar Zr-ZrO$_2$ NPs suggest the oxide grain are almost fully oxidized which is confirmed by our atomic concentration analysis (Fig. 2c), to satisfy the stoichiometric ratio of ZrO$_2$, oxygen can be filled in eight tetrahedral sites (5.5 Å$^3$) of the oxide (Supplementary Fig. 9a); but those tetrahedral sites (4.2 Å$^3$) in Zr metal are too small (Supplementary Fig. 9b). Fig. 2b and Supplementary Movie 3 shows the 3D oxidation maps of Zr1. The degree of oxidation distribution in all the phases are shown in Fig. 2e and Supplementary Fig. 10, where the degree of oxidation increases along with the decrease of Zr $\rho_N$. The c-ZrO$_2$ and a-ZrO$_2$ grains are fully oxidized in their surfaces; and they become less oxidized as closer to their interfaces with the central Zr core (Fig. 2e). The experimentally measured tetrahedral sites in central Zr core are too small to be filled with oxygen, confirming the core is barely oxidized. A cubic cutout of the 3D oxidation maps reveals that the degree of oxidation is strongly correlated to the atomic packing density of Zr; a highly oxidized region always has a lower Zr $\rho_N$ (Fig. 2f). It's notable that some other Zr NPs are completely oxidized to cubic phase ZrO$_2$ and/or amorphous ZrO$_2$ (Supplementary Fig. 11) even from the same batch of oxidation. These results indicate the oxidation process is kinetics controlled, in which we observed several intermediate states of oxidized Zr-ZrO$_2$.

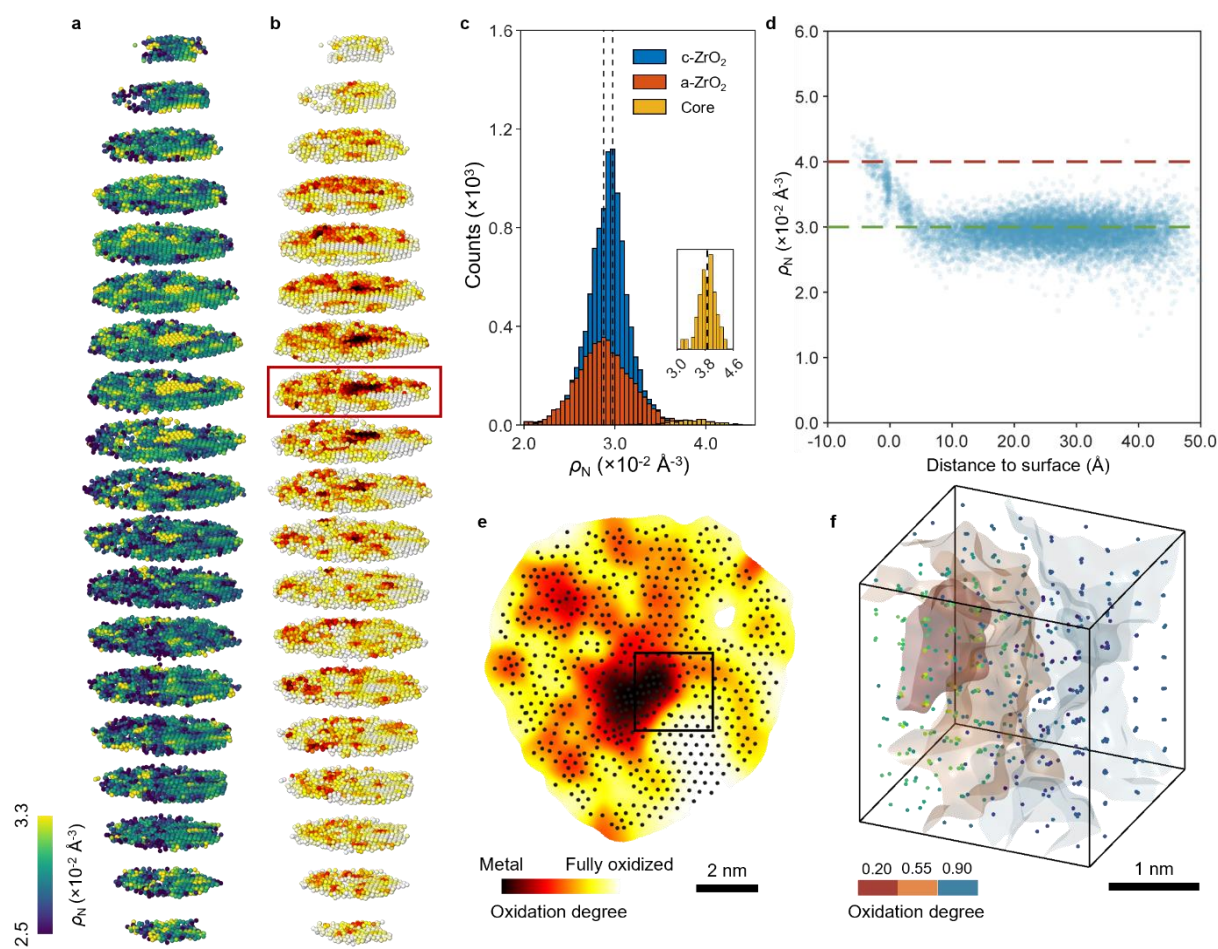

**Fig. 2 | Atomic concentration and the degree of oxidation of Zr-ZrO$_2$ NP. a,b,** Atomic concentration $\rho_N$ distribution (**a**) and the degree of oxidation (**b**) of all the atoms in Zr1. Each slice has a thickness of 5.3 Å. To increase the signal-to-noise ratio, we convolved the degree of oxidation with a 2-Å-wide 3D Gaussian kernel, but this also reduces the 3D spatial resolution of oxidation map to ~4 Å. **c,** Distribution of $\rho_N$ in c-ZrO$_2$ (blue), a-ZrO$_2$ (red) and metal core (yellow) phase. c-ZrO$_2$ has a slightly larger $\rho_N$ distribution by 1% (dashed lines) than a-ZrO$_2$. The inset figure shows the magnified histogram of metal core. **d,** The $\rho_N$ distribution of metal/c-ZrO$_2$ as a function of the distance from the surface of metal core. The dashed lines show the standard $\rho_N$ in Zr metal (red) and cubic phase ZrO$_2$ (green). **e,** A slice through the Zr1 NP as the red rectangle marked in (**b**), showing the degree of oxidation at different regions. **f,** 3D surface rendering of local degree of oxidation and corresponding atomic concentration $\rho_N$, showing the strong correlation. The cutout is 25 × 25 × 25 Å$^3$.

## 3D atomic metal-oxide interfaces

Coherency of the metal-oxide interface affects many properties including strain, diffusion and band structure[26,43,44]. It is extremely difficult to identify the atomic arrangement of semi-coherent or incoherent metal-oxide interfaces from 2D projected images. To probe the 3D structure of metal-oxide interface at atomic level, we focus on the atomic Zr-Zr bonding of the interfaces with a range of ~10 Å based on the packing density between metal core and oxide phases (Fig. 2d). Fig. 3a presents the 3D surface renderings of three major phase, showing the contour of metal core, c-ZrO$_2$ and a-ZrO$_2$ phase. Three slices with four atomic layers in thickness through the metal core show the Zr-Zr bonding of metal-oxide interfaces (Figs. 3b-

3d). We found several types of interfaces including semi-coherent and incoherent interfaces between metal and c-ZrO$_2$, and incoherent interface between metal and a-ZrO$_2$. The white rectangles in Figs. 3b-3d highlight three cutouts from the atomic structures of a semi-coherent (Fig. 3e) and an incoherent interface (Fig. 3k) between metal and c-ZrO$_2$, and an incoherent interface (Fig. 3l) between metal and a-ZrO$_2$, respectively.

In the semi-coherent interface, four layers of metal Zr atoms (marked in deep red) from the metal [1$\bar{1}$0] direction correspond to four layers of Zr atoms (marked as ivory) in the oxide (Fig. 3e). To see the atomic connections in a single corresponding layer, one plane in the cutout is extracted and viewed from [$\bar{1}\bar{1}$1] direction of metal (Fig. 3g, Supplementary Fig. 12). Metal ($\bar{1}\bar{1}$1) plane is almost coplanar with oxide (002) plane; and the interface is about two atomic layers in thickness (blue atoms in Fig. 3g) and primarily connects metal (111) face with oxide (11$\bar{1}$) face. The Zr-Zr bond lengths increase from metal side (~3.3 Å) to oxide side (~3.6 Å). The interface has a long Zr-Zr distance which is due to partially oxidation. Moreover, there is an angular mismatch of ~11° between metal planes and the interfacial planes in metal [112] direction (oxide [110] direction), making the interface bending towards the oxide (Fig. 3e). To better illustrate the origin of the angular mismatch, we build an ideal model of Zr crystal grain and connect it to a cubic phase ZrO$_2$ from the same crystal orientation (Fig. 3h). Since the metal and oxide grains have different crystal orientations, there is a 15° of wedge through direct connection (angular mismatch in Fig. 3h). To minimize the interfacial energy while maintain the coherency, the oxide has to adopt a bending of 15° to fill the wedge (Fig. 3i). At the interface, the maximum numbers of filling oxygen are four instead of eight (Fig. 3j), which means the interface is partially oxidized and the maximum stoichiometric ratio is ZrO. Besides, it is notable that there is a gap angle of ~8° between the Zr (100) planes in the oxide and those in the interface (Fig. 3e), alleviating the overall strain in the whole NPs. By rotating this cutout 120° counter-clockwise, we observed another angular mismatch of ~4° between the metal ($\bar{1}\bar{1}$1) planes and oxide (002) planes in metal [1$\bar{1}$0] direction (oxide [1$\bar{1}$0] direction; Fig. 3f), which is perpendicular to metal [112] direction. It is considerable to have angular mismatch when two adjacent crystal grains having different crystal plane spacing. The (111) spacing of Zr metal is 2.694 Å while (200) spacing of Zr oxide is 2.546 Å. To compensate the spacing mismatch and to maintain the coherency, a certain degree (approximately 4°) of twisting between metal and oxide is preferred (Supplementary Fig. 13)[45].

Most of the metal-oxide interfaces are incoherent in the whole particle. Figs. 3k and 3l show the incoherent interfaces of metal/c-ZrO$_2$ and metal/a-ZrO$_2$, respectively. Although the atomic bonding become more distorted and disordered in the incoherent interfaces between metal and c-ZrO$_2$, most of the metal core {111} faces still correspond to oxide {111} faces (Fig. 3k and Supplementary Fig. 14). Zr atoms form an incoherent boundary with lower coordination number and longer bond length than crystalline region (Supplementary Fig. 15 and Supplementary Fig. 16). Those metal-oxide incoherent interfaces introduce a number of defects which are distributed around the metal core. Many Zr defects are found in those incoherent interfaces. These observations indicate that when semi-coherent interface forms, a significant amount of strain could occur due to lattice and/or angular mismatch during oxidation. Once the strain caused by bending or twisting is too large, some of the semi-coherent interfaces could possibly turn to disordered structures through amorphization[46], where the coherency of interface is completely broken.

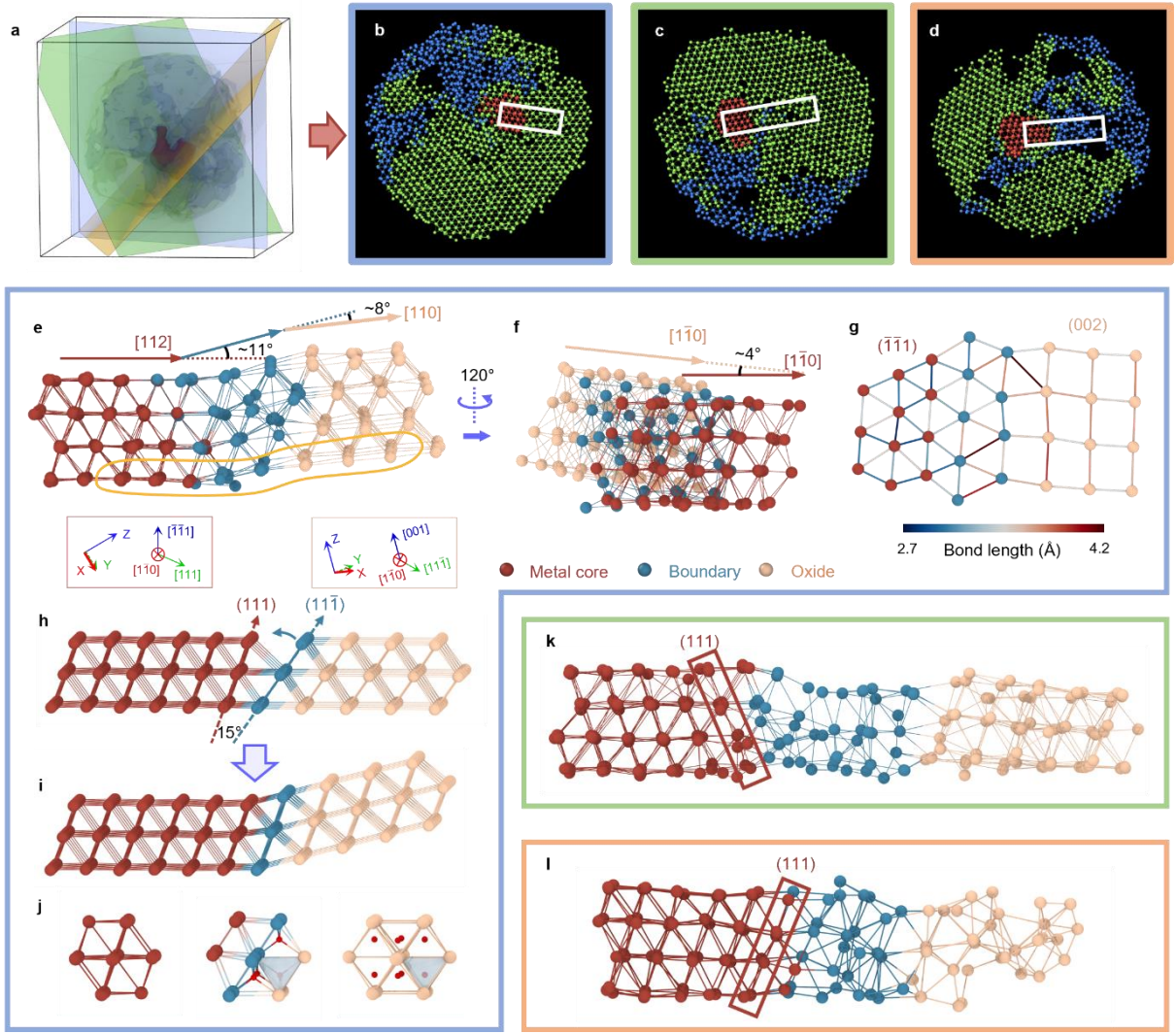

**Fig. 3 | 3D atomic metal-oxide interfaces. a-d**, 3D surface renderings of three major phase, showing the contour (**a**) of metal core (red), c-ZrO$_2$ (green) and a-ZrO$_2$ (blue) of Zr1. Three planes are going through the Zr1 in different directions. The sliced atomic models (4-atom-layers in thickness) highlights three different types of interfaces, i.e., semi-coherent interface between metal core and c-ZrO$_2$ (**b**; in light blue frame), incoherent interface between metal core and c-ZrO$_2$ (**c**; in green frame) and incoherent interface between metal core and a-ZrO$_2$ (**d**; in orange frame). **e-g**, Experimental semi-coherent interface structures specified by the rectangle region in (**b**) and (**h-i**) ideal model built with ideal FCC Zr metal and cubic ZrO$_2$. **e**, The semi-coherent interface viewing from metal [1$\bar{1}$0] direction. There is a bending of ~11° between metal and interfacial layers in metal [112] direction (angle between red line and blue line), and a bending of ~8° between interfacial layers and c-ZrO$_2$ in oxide [110] direction (angle in blue line and ivory line). The coordination tripods in red and ivory boxes shows the spatial crystal orientation of metal and oxide, respectively. **f**, The semi-coherent interface viewing from metal [101] direction (by rotating the cutout in (**e**) 120° counter clockwise), showing a twisting of ~4° in metal [1$\bar{1}$0] direction (angle between red line and ivory line). **g**, One atomic plane extracted from the semi-coherent interface (the highlighted area in red in **e**), viewing from metal [$\bar{1}\bar{1}$1] direction. In this direction, the oxide shows the (002) plane. The color of the atomic bonding shows the Zr-Zr bond length. The Zr-Zr bond lengths in metal and oxide are close to 3.3 Å and 3.6 Å, respectively. The Zr-Zr bond lengths in the interfacial layers are longer. **h**, The ideal model of an interface structure between ideal FCC Zr metal and ideal cubic ZrO$_2$, showing a 15° of wedge if no bending exists, some of the atoms cannot be bonded this way. To minimize the energy and maintain the coherency, a bending of 15° (**i**) is needed to release the stress. The structure changed from metal to oxide shows in (**j**). The oxygen atoms are

colored in red. **k-l**, Incoherent interface structures specified by the rectangle region in (**c**) and (**d**), showing the metal/c-ZrO$_2$ interface (**k**) and metal/a-ZrO$_2$ interface (**l**). In panel **e-l**, the metal atoms, interfacial atoms and oxide atoms are colored in deep red, blue and ivory, respectively. The Zr atoms are bonded with their first-nearest Zr neighbors and linked with lines (Methods).

**Porous structures during oxidation**

Porous structures of the oxide film formed on the surface of metal are usually associated with metal corrosion[11,17,27–29]. We observed numbers of porous structures in the Zr-ZrO$_2$ particles. Fig. 4a shows a 2.4-Å-thick slice from the reconstruction volume of Zr1; in which significant number of voids, such as Zr vacancies (triangle), nano-pores (rectangle) and the largest pore (circle) are observed in the particle. From the 3D intensity and surface renderings of three consecutive atomic layers, a single Zr vacancy defect can be clearly located (Supplementary Fig. 17). To determine all the voids and evaluate their occupied volume, we employed Voronoi analysis by measuring the distance of Voronoi vertices to atoms (Methods). Fig. 4b shows the histogram of volume distribution of all the voids, which we define as Zr vacancies, nano-pores and an extremely large nanoscale pore throughout the particle. The porosity is 17% and 14% in Zr1 and Zr2, respectively. Fig. 4c and Supplementary Movie 4 show the distribution of all Zr vacancies in Zr1. No vacancy is found in the metal core. More than 110 vacancies are distributed in the particle and all the Zr vacancies contribute 8.4% of the total porosity. Slightly more vacancies are found in a-ZrO$_2$ than in c-ZrO$_2$ (Fig. 4d). We plot the density of Zr vacancies from the boundary of the metal core to the surface of the particle (Fig. 4e). Most of the vacancies are distributed in the range of 15 Å between metal core and oxide, which corresponds to the region where Zr packing density exponentially decreases (Fig. 2d). It's notable that we exclude the vacancies from calculating the Zr packing density, the lower $\rho_N$ of interface is independent with the rich vacancies surrounding the metal core. We found 41 nano-pores in the volume range between 125 and 4500 Å$^3$. They are mostly irregular and with a relatively large length-to-radius ratio. Fig. 4f and Supplementary Movie 4 show the distribution of all nano-pores; they mostly sit at the boundaries between c-ZrO$_2$ and a-ZrO$_2$ regions. The largest pore is more than 34000 Å$^3$ and penetrates throughout the whole particle, providing possible pathway for further oxidation (Fig. 4g and Supplementary Movie 4). This pore predominantly sits in the a-ZrO$_2$ regions, connecting and separating all three phases; it terminates at the c-ZrO$_2$ region, releasing a large amount of strain. It's interesting that the Zr-Zr bonds are extremely distorted at the boundary between two c-ZrO$_2$ domains (Figs. 4h and 4i), where some of the Zr atoms turns to be completely amorphous to release strain. Several nano-pores are coincidentally observed at this region near the small amorphous ZrO$_2$ domain. These findings indicate that when the strain reaches to a certain point, possibly higher than the fracture point, the Zr-Zr crystal bonding could turn distorted and amorphous first, and then rupture to form defects to release the strain. It is generally believed that a compact layer of amorphous oxide at the micrometers scale can protect the interior of metal from further oxidation in aluminum[47,48]. While our results reveals that in zirconium oxide, the amorphous oxide regions are substantially more porous than those in the crystalline regions, the voids would further advance the oxidation of Zr metal. We observed variety of voids in Zr NPs at atomic scale to nanometer scale, which are highly related to the structure of incoherent interface. The rearrangements of all the atom positions including distortion, amorphization and the rupture of bonding are possibly due to the massive mass transportation during oxidation at the metal-oxide interfaces facilitated by these voids.

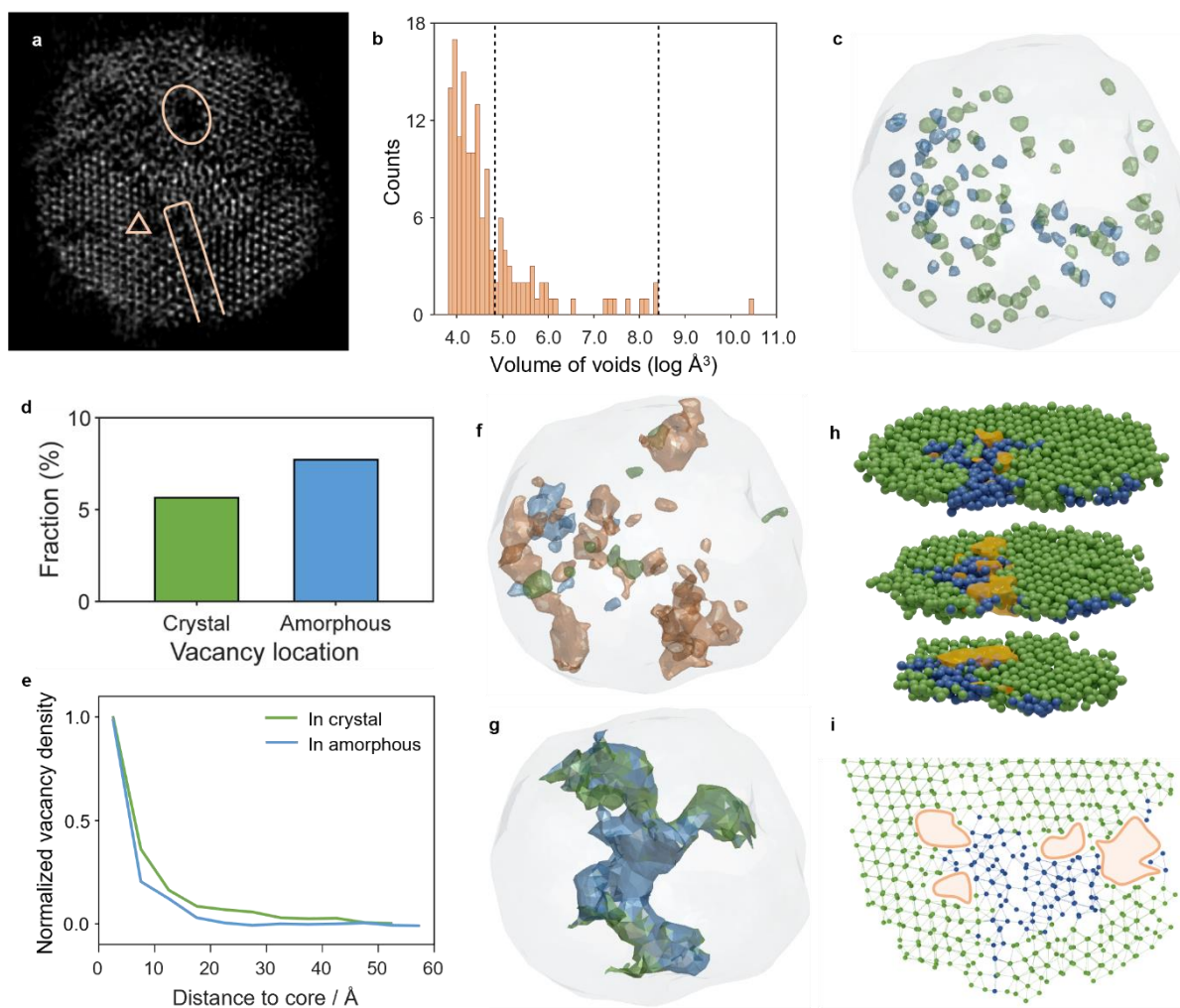

**Fig. 4 | Porous structures during oxidation. a**, A 2.4-Å-thick slice from the reconstructed volume of Zr1, with vacancy (triangle), nano-pores (rectangle) and the largest pore (circle) highlight. **b**, Volume distribution of all the voids. We define the voids with volume no larger than filling two Zr atoms (125 Å$^3$, Methods) as vacancies, the voids with volume between 125 and 4500 Å$^3$ as nano-pores. We consider the largest pore with volume of 34000 Å$^3$ independently as it touches and separates all three phases. Dashed lines show the boundaries between three types of voids we define. **c**, The surface renderings of all vacancies in c-ZrO$_2$ (in green) and a-ZrO$_2$ (in blue). The outline of whole NP is plotted with gray contour. **d-e**, Statistics of vacancies. (**d**) The fractions of vacancies in c-ZrO$_2$ and a-ZrO$_2$. (**e**) The radially normalized density distribution of vacancies as a function of distance from the surface core to the surface. **f**, The surface renderings of all nano-pores in c-ZrO$_2$ (in green), in a-ZrO$_2$ (in blue) and in between c-ZrO$_2$ and a-ZrO$_2$ (in orange). **g**, The surface rendering of the largest pore. The boundary atoms composed of amorphous and crystalline atoms are colored by blue and green, respectively. **h**, One interface between two c-ZrO$_2$ regions with distorted interfacial Zr-Zr bonds, amorphous region and nano-pores. The crystal and amorphous atoms distinguished by BOO analysis are colored as green and blue, respectively. The contour of the nano-pore is colored as orange. **i**, Three representative slices show a 7.8-Å-thick (approximately five atomic layers) cross section of the nano-pores and surrounding atoms.

## Conclusion

In conclusion, we determined the 3D atomic structure of metal-oxide interfaces in Zr-ZrO$_2$ NP for the first time using atomic resolution electron tomography. We quantitatively measured the atomic packing density and the degree of oxidation from our experimental model of metal-oxide interface. The degree of oxidation from metal to oxide increases gradually, resulting a diffuse interface between FCC Zr core and ZrO$_2$. The Zr metal connects with its oxide via {111} planes; and the semi-coherent interface has severe distortion including bending and twisting. The significant stress in the interface is relieved through low coordination and defects. Numbers of defects including vacancies and nano-pores together leverage the mass transportation during oxidation. We anticipate that our new findings will fulfill the dearth of 3D atomic structure of metal-oxide interface and advance the study of fundamental problems of metal-oxide interfaces such as oxidation kinetics, diffusion and defect evolution in variety of materials.

**Methods**

**Synthesis of zirconium nanoparticles**

The Zr NPs were synthesized in liquid using laser ablation methods. An all-solid-state ultraviolet laser with a wavelength of 355 nm was employed for laser ablation in ethanol, with a pulse width of 7 ps, a max pulse energy of 18 μJ, a repetition frequency of 800 kHz, and a beam spot diameter of 20 μm. Before being placed in a clean beaker, the bulk Zr target (purity >99.95%) was washed by acetone (99.5%) and ethanol (99.9%). The dissolved oxygen in liquid is eliminated to the minimization by nitrogen flow for 60 min with a flow rate of 4 L/min. Then, the Zr target was submerged into ethanol and the laser beam was accurately focused vertically on the surface of the bulk Zr through the ethanol in a closed chamber. The laser ablation was continued for 1 min, and the produced Zr NPs were dispersed using ultrasonic agitation and subsequently isolated via centrifugation to be collected in the ethanol solution. The freshly-prepared Zr NPs were then placed in air for one month to obtain a naturally oxidized layer several nanometers thick. The detailed methods of synthesis are described elsewhere.[39] The final Zr-ZrO$_2$ NPs were drop cast onto 7-nm-thick Si$_3$N$_4$ membranes using atomizer for TEM experiment.

**Atomic-resolution electron tomography**

EELS maps were collected on an aberration-corrected Thermo Fisher Scientific Spectra 300 microscope operated at 300 kV using a Gatan Continuum GIF with a K3 direct electron detector in the Bay Area Centre for Electron Microscopy at Songshan Lake Materials Laboratory. EDS maps were collected on an aberration-corrected Thermo Fisher Scientific Themis Z microscope at 300 kV with a 50 pA beam current and a total acquisition time of approximately 5 min in Analytical Instrumentation Center at Peking University.

Tomographic tilt series are acquired by Thermo Fisher Scientific Titan microscope with spherical aberration correction at Electron Microscopy Laboratory of Peking University. The acceleration voltage was 300 kV and the imaging mode was HAADF mode. The tomographic tilt series were acquired at very low dose rate ($< 5 \times 10^5$ e/Å$^2$) to protect the structure of oxide. For each tilt angle, three sequential images with a dwell time of 2 to 4 μs were acquired and registered using normalized cross-correlation, and then the averaged to enhance the signal-to-noise ratio.

Acquired images were drift corrected, denoised and aligned before reconstruction. Linear drift from the sample or stage was corrected during the image registration. Block-matching and 3D filtering (BM3D) is employed to denoise the images after drift correction[49]. And then, the background was estimated using the discrete Laplacian function of MATLAB and subtracted. In the direction perpendicular to the tilt axis, the images were aligned by maximizing the cross-correlation between the common lines. Along the tilt axis, the images were aligned using the center-of-mass method.

After image processing, the 3D reconstruction was computed from experimental tilt series using Real Space Iterative Reconstruction (RESIRE) algorithm[33].

After reconstruction, atom tracing was performed to determine the 3D atomic coordinates. First, we interpolated reconstructed volume with spline method. All the local maxima in the reconstruction were identified as the rough atomic coordinates. Then, the coordinates were optimized according to the local volume of 1.7 Å × 1.7 Å × 1.7 Å with a polynomial fitting method. To separate the non-atoms from the potential atoms, K-means clustering method was employed based on the integrated intensity of the local volume (1.7 Å × 1.7 Å × 1.7 Å). For every potential atom, a minimum distance of 2 Å to its nearest atom should be satisfied. By carefully comparing the individual atom in the potential atomic models with the reconstructed volume, we manually corrected the atomic coordinates of unidentified or misidentified atoms (typically < 1%). The more detailed atom tracing procedure is described elsewhere.[33]

**Calculation of PDF**

We calculated the PDF curve from experimental 3D atomic model by

$$g(r) = \frac{1}{N^2} \sum_{i=1}^{N} \sum_{j=1}^{N} \langle \delta(|\mathbf{r}_{ij}| - r) \rangle$$

where $N$ is the total number of atoms; $\delta$ is the Dirac delta function; $\langle \cdot \rangle$ is the notation for expectation; $|\mathbf{r}_{ij}|$ is the distance between the $i$-th atom and the $j$-th atom. To get a smoother PDF curve, a Gaussian kernel function with a $\sigma$ of 1.5 Å was applied to convolute with original $g(r)$. Finally, the PDF was scaled to approach one at the large pair distances. Using this procedure, we calculate the c-$ZrO_2$ and a-$ZrO_2$ separately in three NPs. From the PDF, we determined the first valley position as 4.5 Å, corresponding to the first-nearest-neighbor shell distance. The distance was used as a cutoff for BOO and alpha shape calculation (see the sections below).

**Local BOO parameters**

We calculated the normalized local BOO parameters to distinguish the order and disorder of all the atoms. The normalized BOO parameter is defined as $\sqrt{\bar{Q}_4^2+\bar{Q}_6^2}/\sqrt{\bar{Q}_{4\,\text{FCC}}^2+\bar{Q}_{6\,\text{FCC}}^2}$, where the $\bar{Q}_4$ and $\bar{Q}_6$ values were computed based on the procedure described elsewhere, using 4.5 Å (the first-nearest-neighbor shell distance) as a constraint[50]. The $\bar{Q}_{4\,\text{FCC}}$ and $\bar{Q}_{6\,\text{FCC}}$ are the reference values of the standard FCC structure. We separated the amorphous part from crystalline part according to the criterion of the normalized BOO less than 0.5[34].

**Determination of voids**

Delaunay triangulation and Voronoi tessellation were performed to determine the voids. Delaunay triangulation, Voronoi tessellation and alpha shape were performed with the built-in functions of MATLAB (namely, 'delaunayTriangulation', 'voronoin', and 'alphaShape'). The initial spatial region of NP was calculated by alpha shape with $\alpha = 4.5$ Å (the first-nearest-neighbor shell distance). Then, we calculated the space that accommodates at least one Zr atom in the initial particle region with the following steps:

(1) The initial particle region was divided into tetrahedra by Delaunay triangulation.

(2) We determined whether a tetrahedron is void. We calculated the radius of circumscribed sphere for each tetrahedron. The radius represents the maximized sphere that can fit within the NP without intersecting with the center of any atom. Tetrahedra with a radius larger than 3.19 Å were classified as voids. This criterion of radius was obtained based on the standard cubic phase of $ZrO_2$ with one vacancy.

(3) We grouped neighboring voids together to form larger voids. Two voids that share a common face are considered neighboring and thus combined into a single, larger void. The volume of these larger voids was calculated by summing the volumes of each void of component.

(4) We classified the voids into vacancies, nano-pores and the largest pore based on volume. We define the voids with volume filling one or two Zr atoms (45-125 Å$^3$) as vacancies, the voids with volume between 125 and 4500 Å$^3$ as nano-pores. The largest pore with volume of 34000 Å$^3$ was considered independently.

Finally, the contours of voids were displayed with Laplacian or HC smoothing conducted by GIBBON[51].

**Atomic concentration**

Topological bonds were determined based on the Voronoi tessellation. Two atoms are considered topologically bonded if their corresponding Voronoi polygons share a common face. In constructing the Voronoi polygons, we removed those surfaces with area less than 1% of the total area of the polygon surfaces[52]. Additionally, this bond must also be shorter than 4.5 Å, corresponding to the first-nearest-neighbor shell distance. The atomic concentration was calculated by $\rho_N = 1/V$, where $V$ is the volume of Voronoi cell of an atom.

**The degree of oxidation**

The oxidation state was determined using Delaunay triangulation. First, the distortion of Delaunay tetrahedra was considered. The distortion parameter was calculated by $\delta = e_{max}/e_{avg} - 1$, where $e_{max}$ and $e_{avg}$ are the maximum and average edge lengths of tetrahedron[33]. We removed the tetrahedron with a distortion parameter larger than 0.255. Then, the volume of remaining Delaunay tetrahedra was calculated. If the volume of a tetrahedron is larger than 4.68 Å$^3$ (the averaged tetrahedron volume of FCC Zr lattice and c-ZrO$_2$ lattice), an oxygen atom was placed inside. Finally, the degree of oxidation for each Zr atom was calculated by the fraction of its surrounding tetrahedra that accommodate one oxygen atom.

**Acknowledgments:** We thank the support of High-performance Computing Platform of Peking University. We thank the Electron Microscopy Laboratory at Peking University, Bay Area Centre for Electron Microscopy at Songshan Lake Materials Laboratory and Analytical Instrumentation Center at Peking University for the use of the aberration-corrected electron microscope. This work was supported by the National Natural Science Foundation of China (Grant No. 22172003, 52071222) and Guangdong Major Project of Basic and Applied Basic Research, China (Grant No. 2019B030302010).

**Author contributions:**

J.Z. conceived the idea and directed the study. Z.L., Z.X. and Y.Z. performed TEM experiment and acquired the data. Y.Z. and S.H. performed the imaging processing, reconstructions, and atom tracing. Y.Z., Z.L., S.H. conducted/discussed data analysis under the direction of J.Z.. T.X. and Y.-E. Z. synthesized Zr NPs under the direction of H.-B. K. and W.-H. W. Y.Z., Z.L. and J.Z. wrote the manuscript. All authors commented on the manuscript.

**Competing interests:** The authors declare no competing interests.

**Data availability:** All data are available upon reasonable request.

# Supplementary Information for

# Three-dimensional atomic interface between metal and oxide in Zr-ZrO$_2$ nanoparticles

Yao Zhang[1,5], Zezhou Li[1,5], Xing Tong[2,5], Zhiheng Xie[1], Siwei Huang[1], Yue-E Zhang[2,3], Hai-Bo Ke[2]*, Wei-Hua Wang[2,4], Jihan Zhou[1]*

[1]*Beijing National Laboratory for Molecular Sciences, Center for Integrated Spectroscopy, College of Chemistry and Molecular Engineering, Peking University; Beijing, 100871, China.*
[2]*Songshan Lake Materials Laboratory, Dongguan 523808, China.*
[3]*College of Physics, Liaoning University, Shenyang 110036, China.*
[4]*Institute of Physics, Chinese Academy of Sciences, Beijing 100190, China.*
[5]*These authors contributed equally to this work.*
*\*Correspondence and requests for materials should be addressed to H.-B. K. (email: kehaibo@sslab.org.cn) and J. Z. (email: jhzhou@pku.edu.cn)*

**Supplementary Table 1** Tomography and reconstruction overview

|  | Zr1 | Zr2 | Zr3 |
|---|---|---|---|
| **Data Collection and Processing** |  |  |  |
| Voltage (kV) | 300 | 300 | 300 |
| Convergence semi-angle (mrad) | 21.4 | 21.4 | 21.4 |
| Detector inner angle (mrad) | 39.4 | 31.5 | 31.5 |
| Detector outer angle (mrad) | 200 | 190.6 | 190.6 |
| Pixel size (Å) | 0.343 | 0.343 | 0.343 |
| Scanning current (pA) | 15 | 15 | 15 |
| Number of projections | 50 | 59 | 57 |
| Tilt range (°) | -75 to 76 | -76 to 77 | -75 to 75.5 |
| Electron dose ($10^5$ e·Å$^{-2}$) | 4.8 | 3.4 | 2.9 |
| **Reconstruction** |  |  |  |
| Algorithm | RESIRE | | |
| Oversampling ratio | 4 | 4 | 4 |
| Number of iterations | 200 | 200 | 200 |
| Reconstruction error (%) | 0.1069 | 0.0728 | 0.0990 |
| Number of atoms | 15392 | 22382 | 13482 |

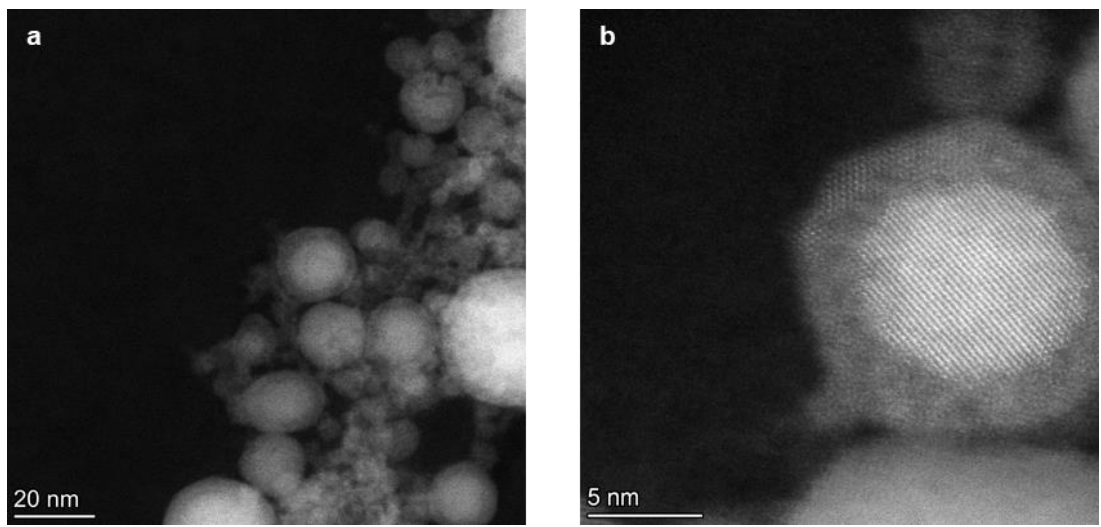

**Supplementary Figure 1** HAADF-STEM images show the Zr-ZrO$_2$ NPs with an oxidized shell and a metal core. **a**, A relatively low magnification image shows overview of the oxidized Zr NPs. **b**, A high-resolution image shows a representative Zr-ZrO$_2$ NP with a metal core (high intensity) and an oxide shell (low intensity).

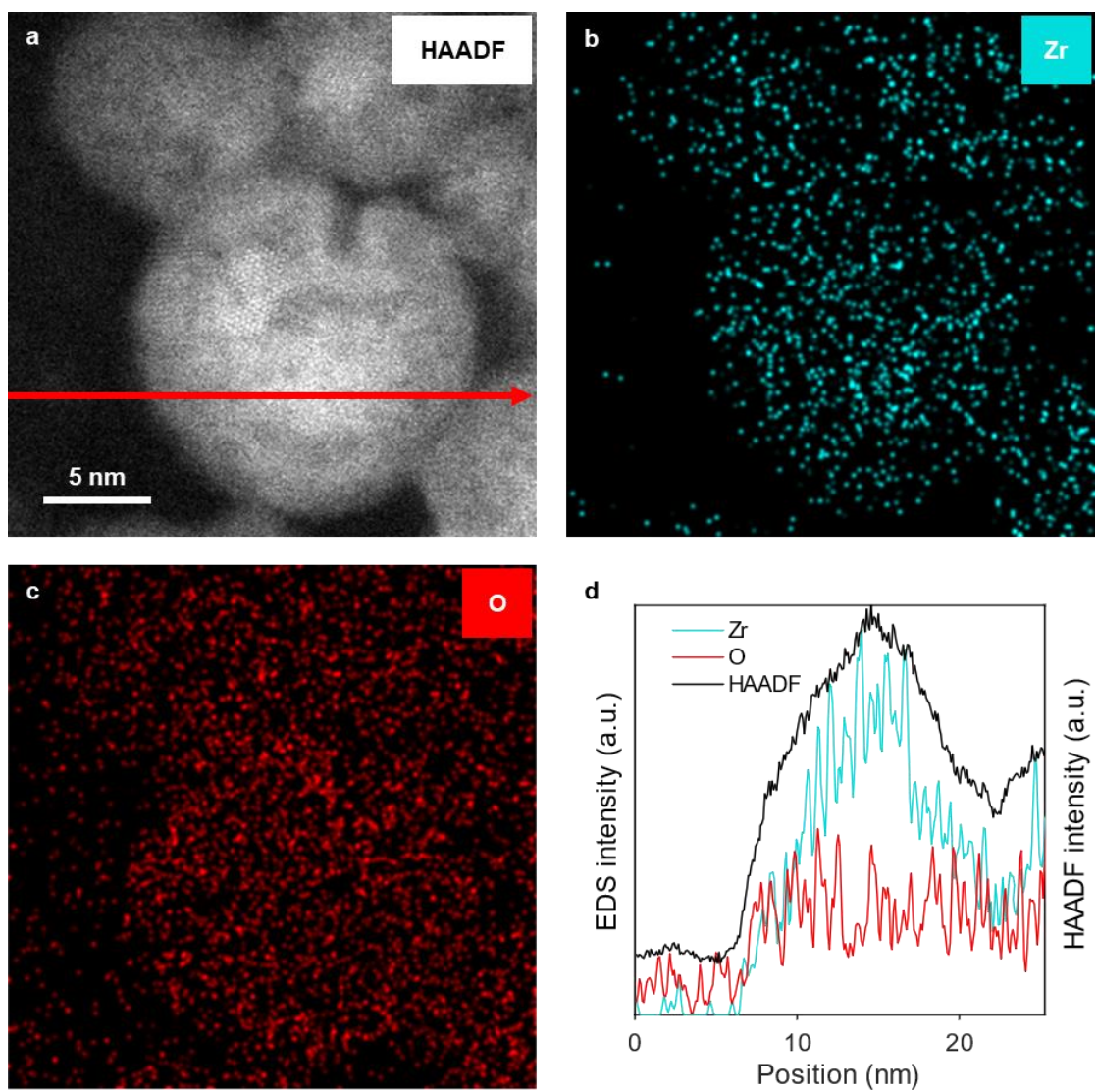

**Supplementary Figure 2** EDS mapping and profile show the elements distribution of Zr-$ZrO_2$ NPs. The HAADF-STEM image (**a**) and its corresponding Zr (**b**), O (**c**) EDS mapping, with high Zr signals and low O signals at the center of the NP confirmed by the line profile (**d**) along the red arrow in (**a**).

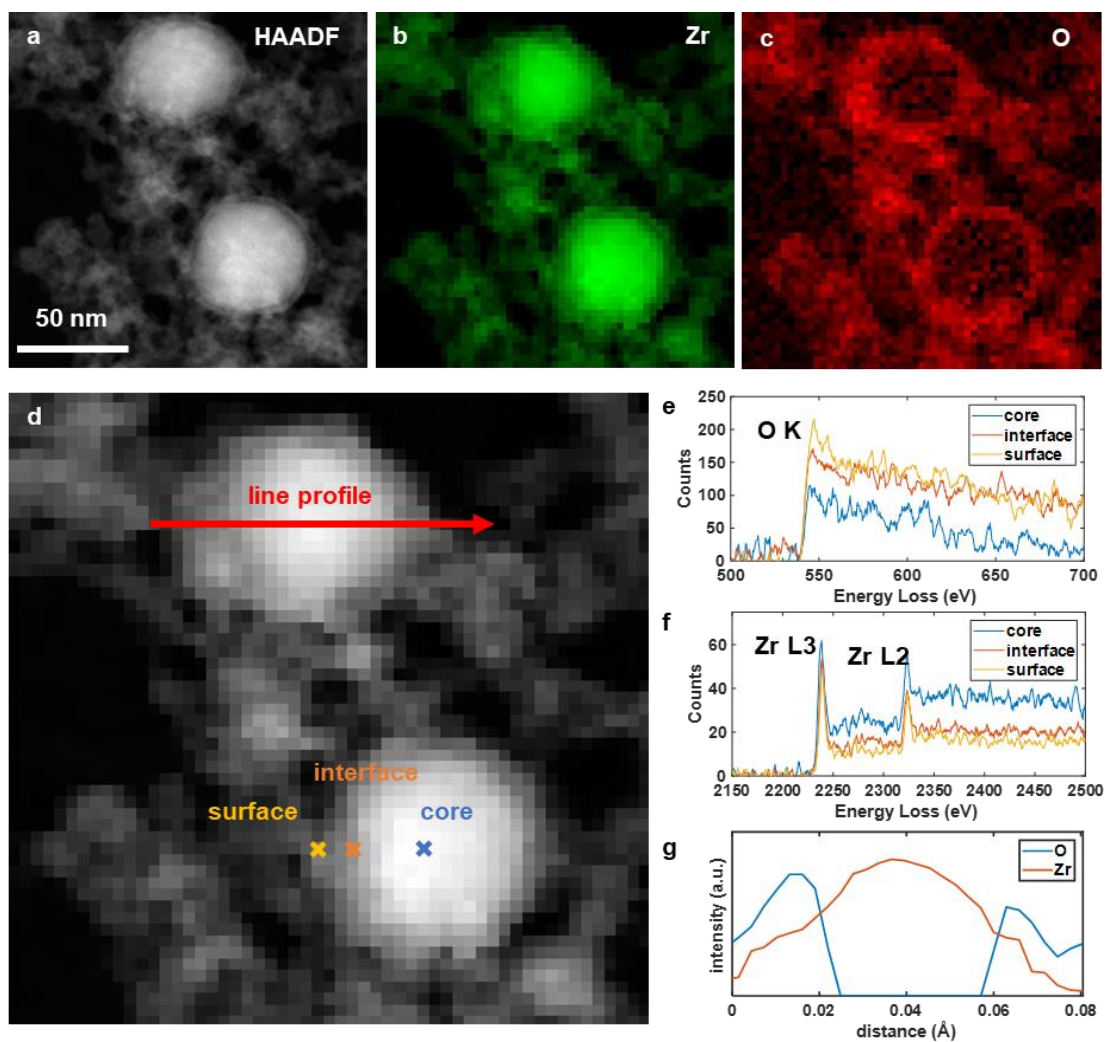

**Supplementary Figure 3** EELS mapping shows the elements distribution of Zr-ZrO$_2$ NPs. The HAADF-STEM image (**a**) and its corresponding Zr (**b**), O (**c**) EELS mapping, with high Zr signals at the center of the NP and high O signals at the edge. The EELS spectrums of O (**e**) and Zr (**f**) from the points picked in (**d**) at core (blue), interface (orange) and surface (yellow). (**g**) The intensity profile of Zr and O respectively along the red arrow in (**d**).

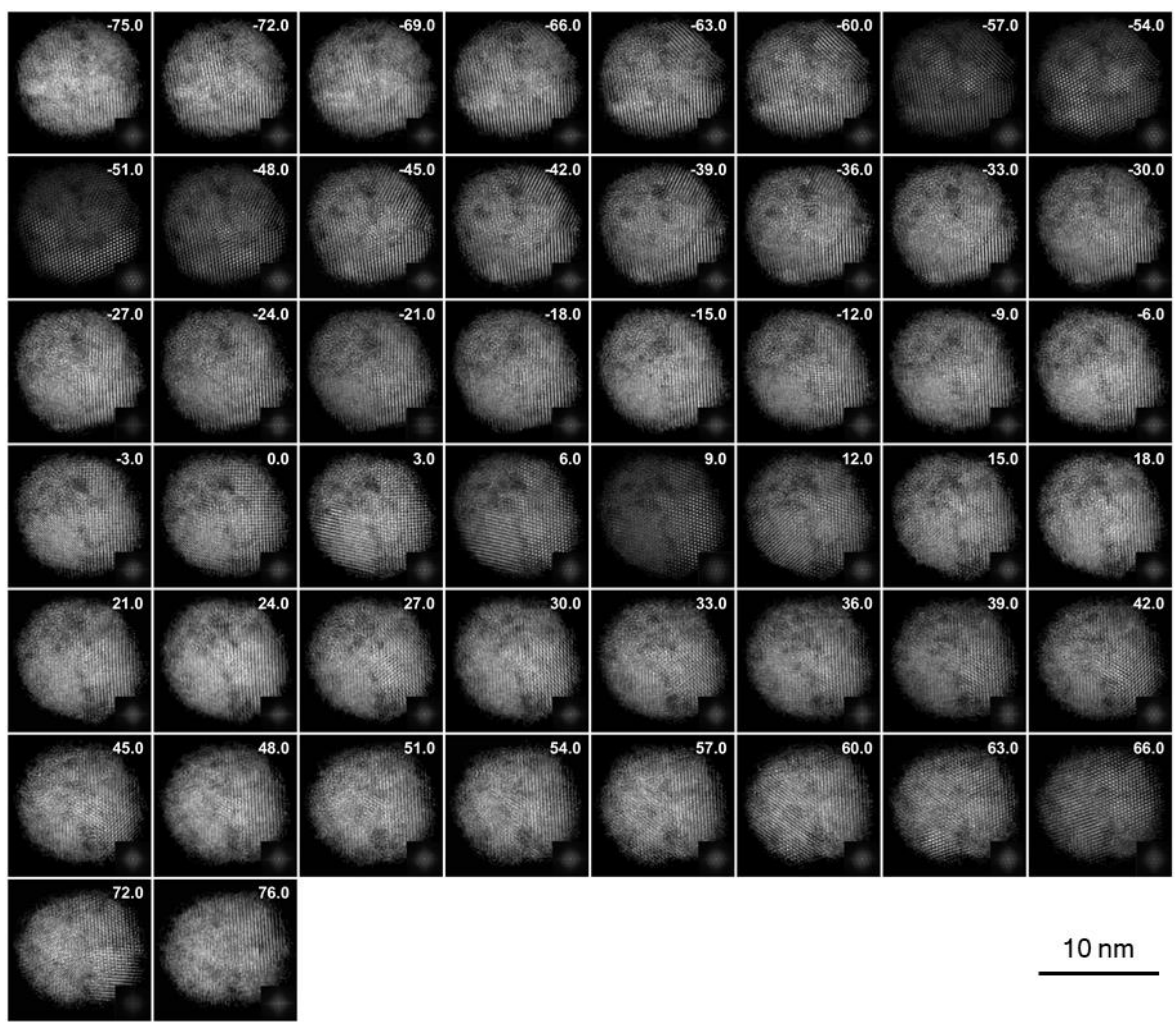

**Supplementary Figure 4** Tomography tilt series of the Zr1 NP. 50 HAADF-STEM images with a tilt range from −75.0° to +76.0°. The Fourier transform of each image is displayed in the bottom-right corner.

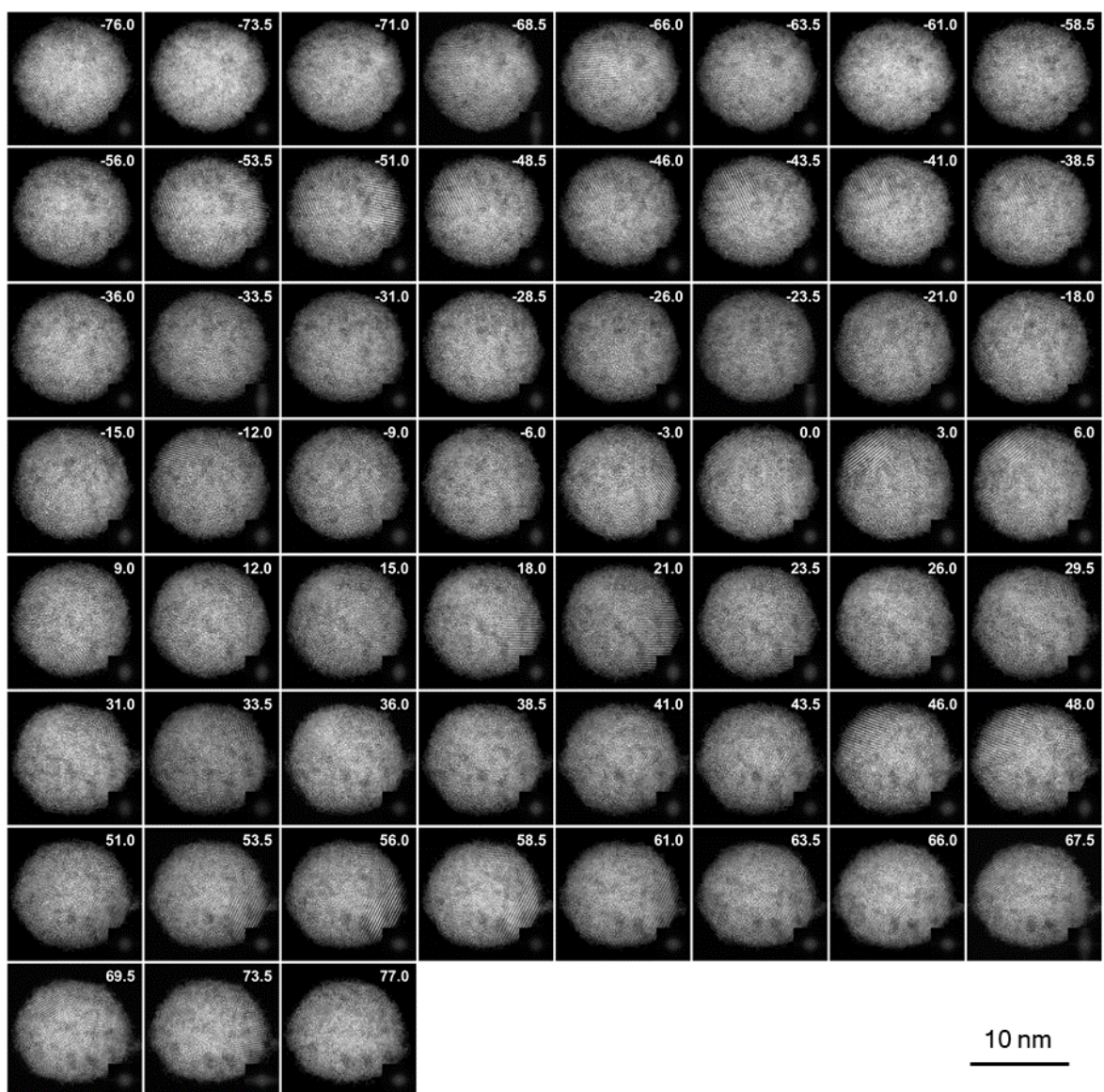

**Supplementary Figure 5** Tomography tilt series of the Zr2 NP. 59 HAADF-STEM images with a tilt range from −76.0° to +77.0°. The Fourier transform of each image is displayed in the bottom-right corner.

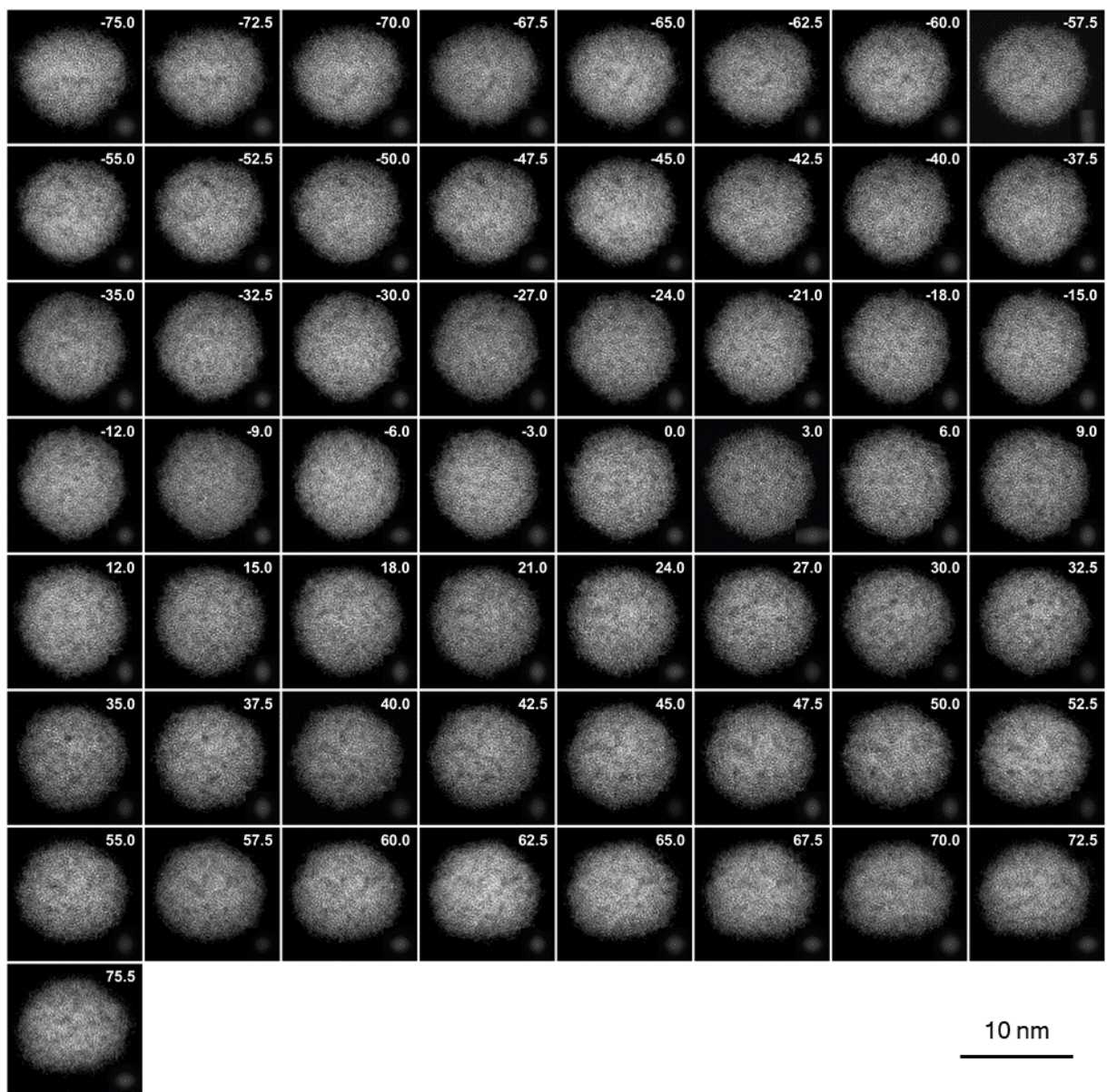

**Supplementary Figure 6** Tomography tilt series of the Zr3 NP. 57 HAADF-STEM images with a tilt range from −75.0° to +75.5°. The Fourier transform of each image is displayed in the bottom-right corner.

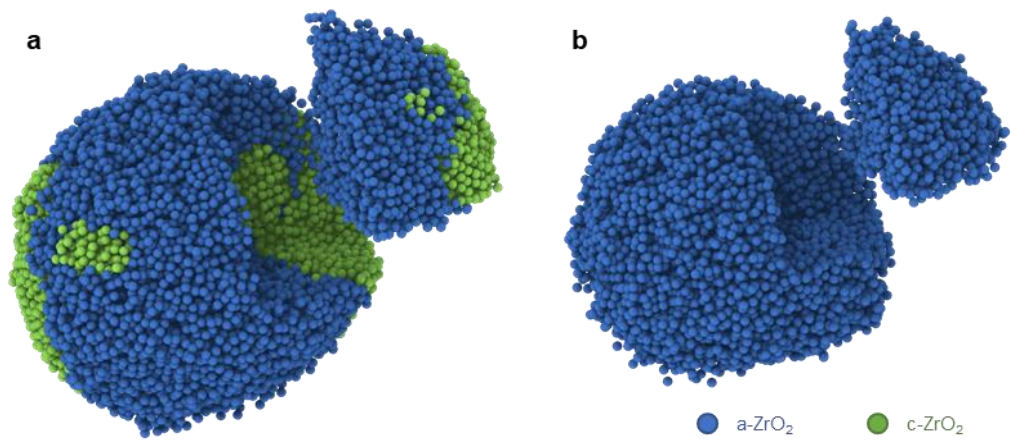

**Supplementary Figure 7** Atomic model for Zr2 and Zr3 NPs. **a**, Atomic model for Zr2 NP, with crystalline oxide grains (c-ZrO$_2$; in green) and a large amorphous oxide phase (a-ZrO$_2$; in blue). **b**, Atomic model for Zr3 NP, which contains a-ZrO$_2$ only.

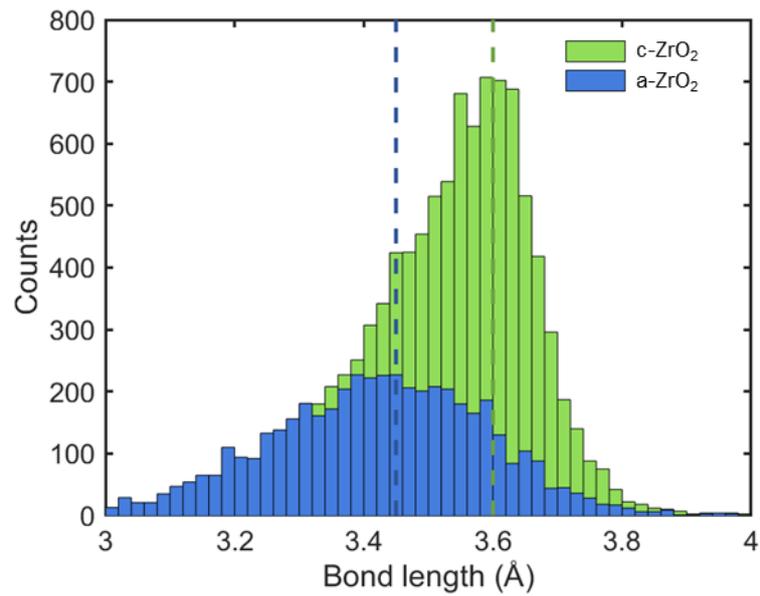

**Supplementary Figure 8** Bond length distribution of Zr1 NP, with c-ZrO$_2$ in green and a-ZrO$_2$ in blue. The histogram shows the distribution of bond length in c-ZrO$_2$ and a-ZrO$_2$. The dashed lines show the most populated bond length in c-ZrO$_2$ (3.6 Å; the green line) and a-ZrO$_2$ (3.45 Å; the blue line).

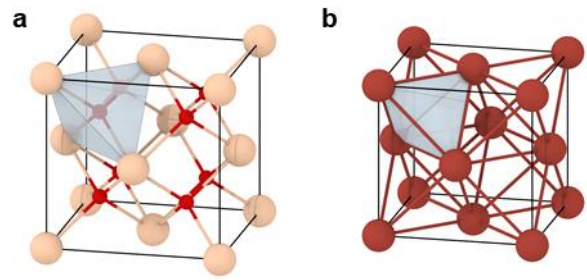

**Supplementary Figure 9** The volume difference of the tetrahedral site in pure FCC Zr and c-ZrO$_2$. **a**, The tetrahedral site in c-ZrO$_2$, which accommodates one oxygen atom. The atoms in ivory are oxide Zr atoms, and the atoms in red are O atoms. **b**, The tetrahedral site in FCC Zr. The atoms in deep red are metal Zr atoms. Both in (**a**) and (**b**), the tetrahedron is displayed with surface rendering in light blue.

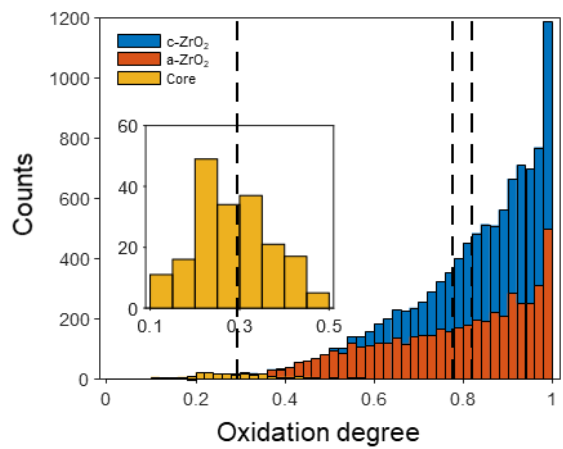

**Supplementary Figure 10** The distribution of degree of oxidation. The histogram shows the distribution of degree of oxidation in c-ZrO$_2$, a-ZrO$_2$ and metal core. The dashed lines are the mean values of degree of oxidation for c-ZrO$_2$ (0.82), a-ZrO$_2$ (0.78) and metal core (0.29) from right to left. The inset figure shows the magnified histogram of metal core.

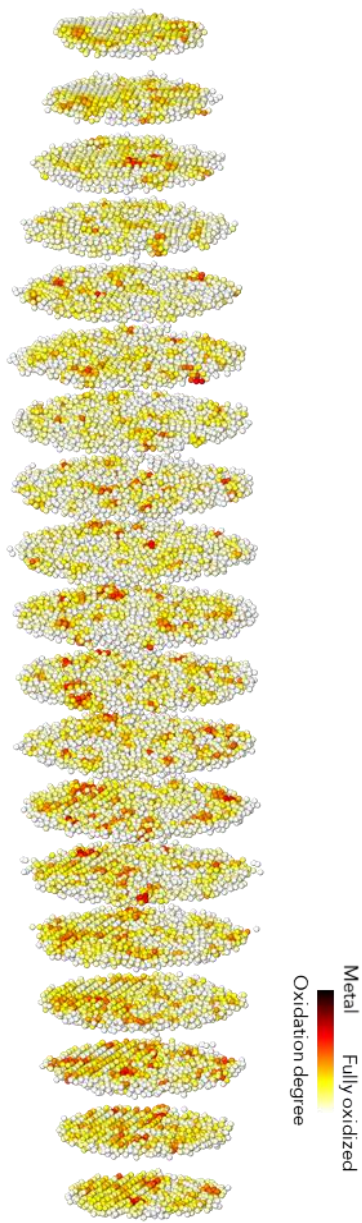

**Supplementary Figure 11** Atoms of the Zr2 NP colored according to the degree of oxidation, divided into slices with a thickness of 5.3 Å. The degree of oxidation shows that the Zr2 NP is completely oxidized compared with Zr1 NP.

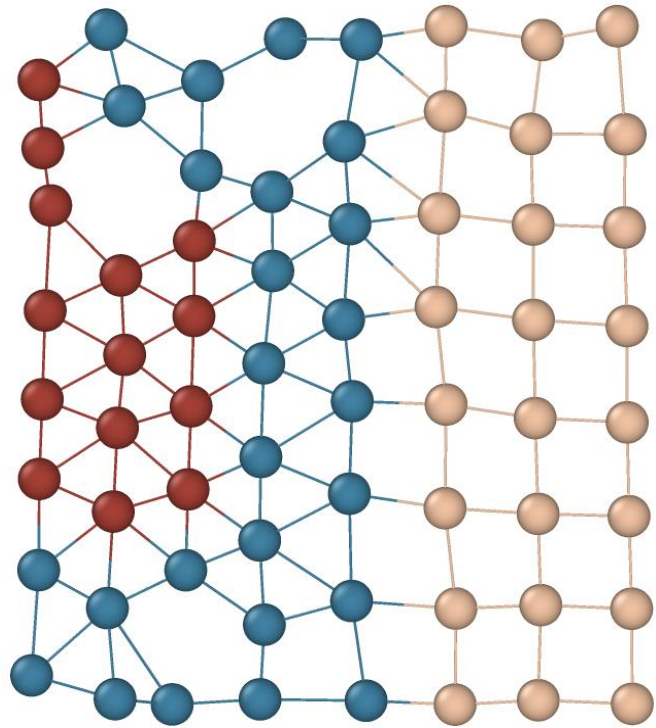

**Supplementary Figure 12** The surroundings of the plane in Figure 3g, with the dislocations and vacancies caused by semi-coherent interface. The metal atoms, interfacial atoms and oxide atoms are colored in deep red, blue and ivory, respectively.

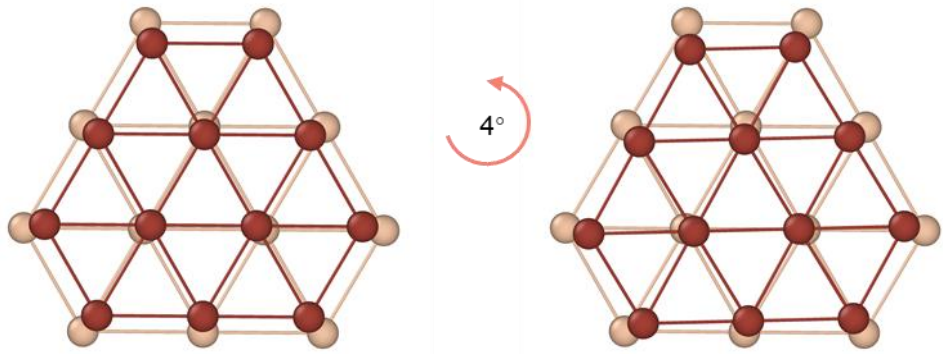

**Supplementary Figure 13** The model for the twisting. This figure shows the corresponding {111} planes of metal core and c-ZrO$_2$ connect with each other. This twisting minimizes the energy of semi-coherent interfaces. The metal Zr atoms and oxide Zr atoms are colored in deep red and ivory, respectively.

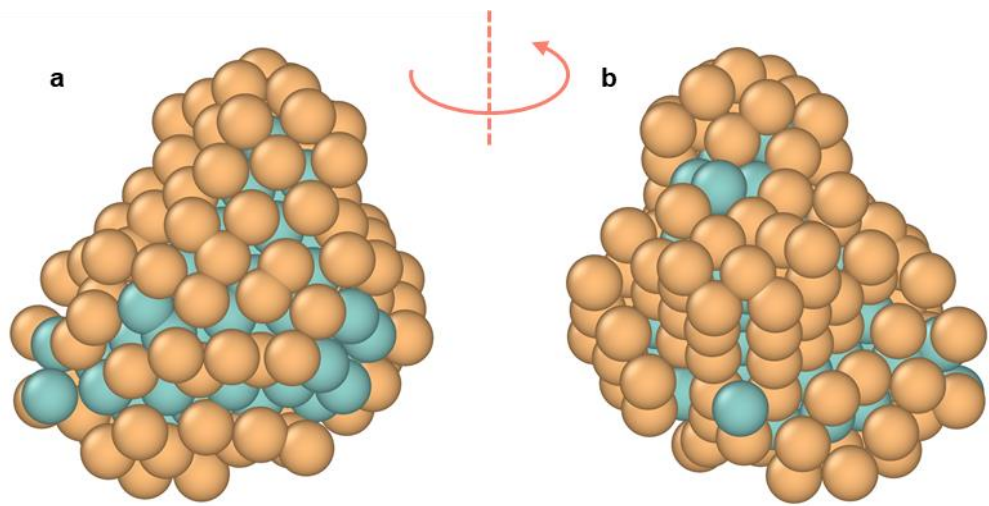

**Supplementary Figure 14** All {111} faces are colored in orange in metal core, while the other atoms are colored in green. **a**, In the direction of one {111} face. **b**, Rotate the metal core to the back.

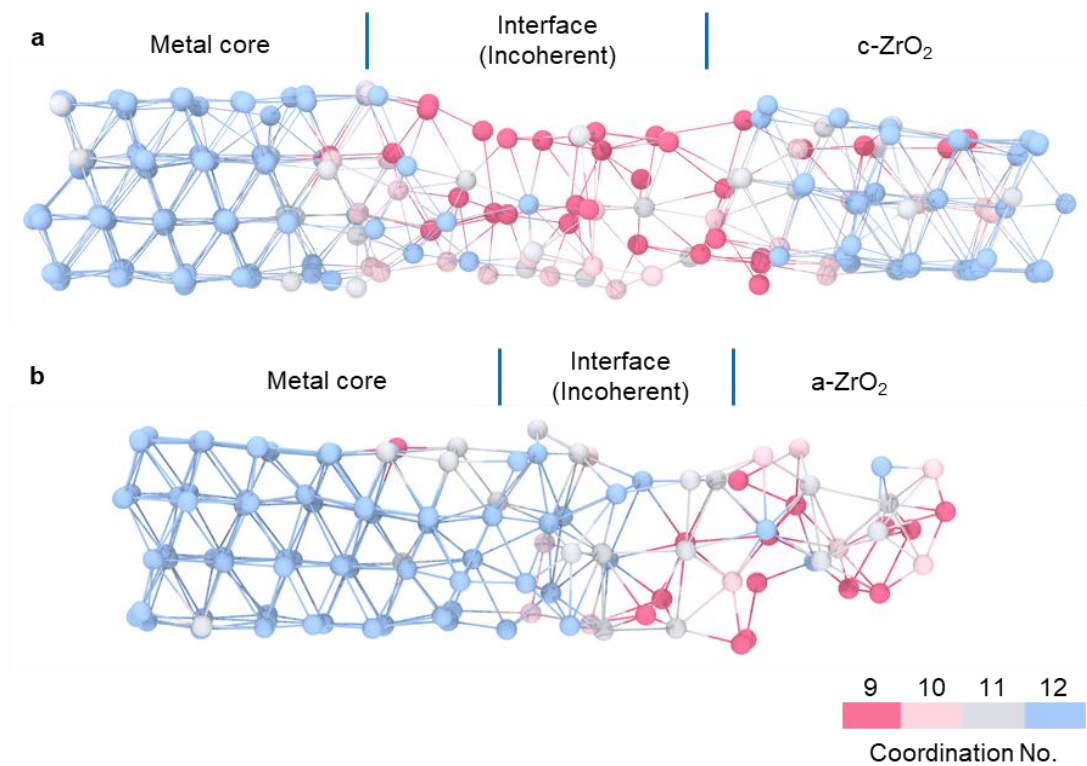

**Supplementary Figure 15** Coordination number of metal/c-ZrO$_2$ (**a**) and metal/a-ZrO$_2$ (**b**) incoherent interface. The coordination number is lower at the incoherent interface.

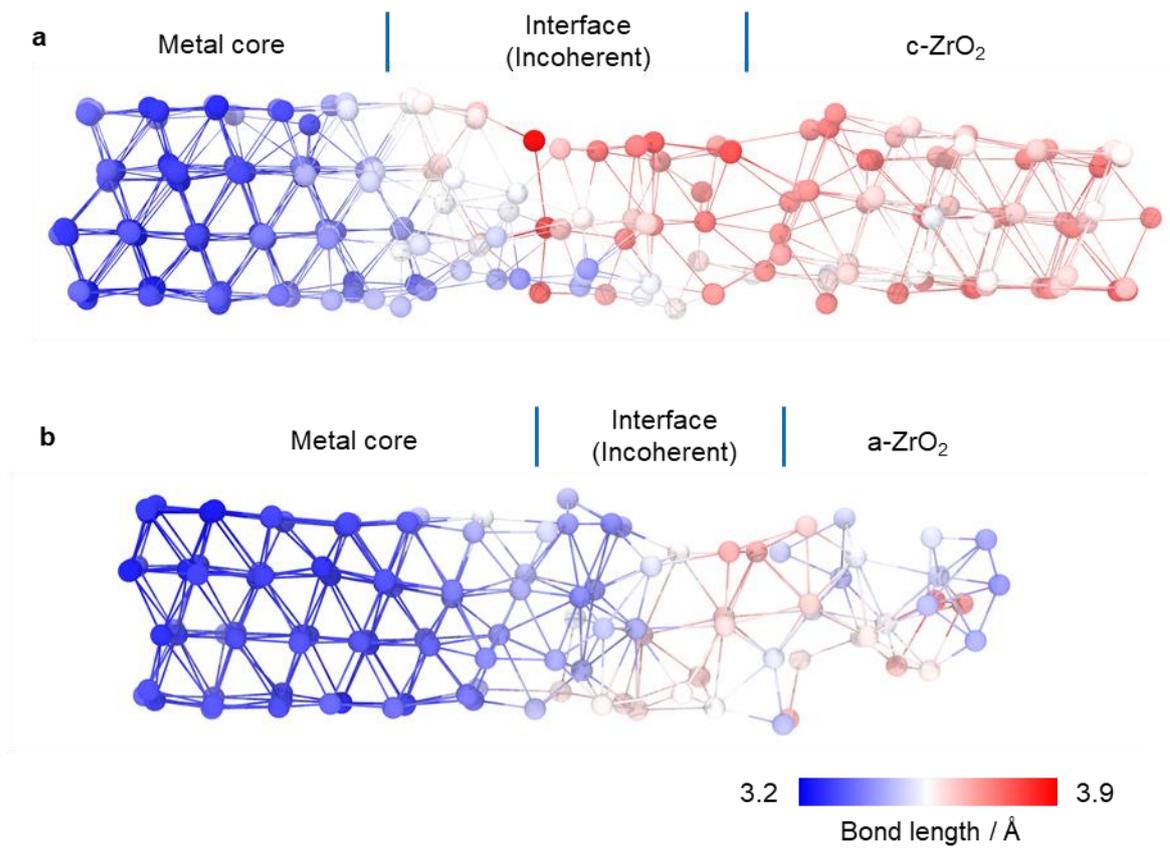

**Supplementary Figure 16** Averaged bond length (in Å) of metal/c-ZrO$_2$ (**a**) and metal/a-ZrO$_2$ (**b**) incoherent interface. The bonds are longer at the incoherent interface.

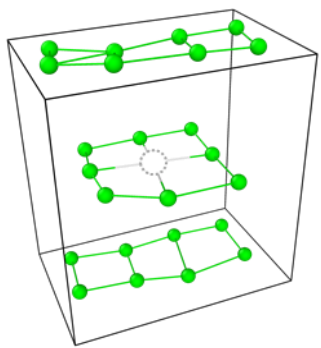

**Supplementary Figure 17** A Zr vacancy displayed with a dashed circle in Zr1 NP. The Zr atoms are colored in green.